\documentclass[twocolumn,twoside,slac_two]{revtex4}
\usepackage{graphicx}
\usepackage{fancyhdr}

\begin{document}

\title{Low scale leptogenesis with an $SU(2)_L$ singlet scalar}

\author{M. Frigerio}
\affiliation{Service de Physique Th\'{e}orique, CEA/Saclay, F-91191 Gif-sur-Yvette Cedex, FRANCE}

\begin{abstract}
We consider thermal leptogenesis as the origin of 
the matter-antimatter asymmetry of the Universe.
Some phenomenologically unpleasant features of the usual leptogenesis scenario are reviewed. We propose a minimal alternative,  which makes use of a charged scalar 
$\delta^+$, singlet under $SU(2)_L$. Such particle is contained in natural extensions of the Standard Model. The lepton asymmetry is generated by the decays of a right-handed neutrino into $\delta^+$ and a right-handed charged lepton
and it turns out to be sufficiently large for scales as low as $\sim$ TeV. The phenomenological signatures at colliders are discussed.
\flushright{SACLAY-T06/125}
\end{abstract}

\maketitle

\thispagestyle{fancy}


\section{LEPTOGENESIS: MOTIVATIONS AND SOME DRAWBACKS}

A very well motivated explanation for non-zero tiny neutrino mass is 
lepton number violation (LNV) at an energy scale $M$ much larger than the electroweak symmetry breaking scale $v$. This leads to
a Majorana-type neutrino mass  $m_\nu\sim v^2/M$.
The most popular way to introduce LNV is to add to the Standard Model (SM) sterile right-handed (RH) neutrinos $N_i$, with large Majorana masses (seesaw mechanism).
These sterile neutrinos decay in the early Universe, at a temperature of the order of their mass. If the decays occur (partially) out-of-equilibrium and violate CP, a different number of leptons and antileptons is produced. Such lepton asymmetry is efficiently converted into a baryon asymmetry by SM $(B+L)$-violating effects (sphalerons), which are in thermal equilibrium down to temperatures $T \sim v$.
This mechanism to explain the matter-antimatter asymmetry of the Universe is known as baryogenesis via leptogenesis \cite{fy}.

\begin{figure*}[t]
\centering
\includegraphics[width=30mm,angle=270]{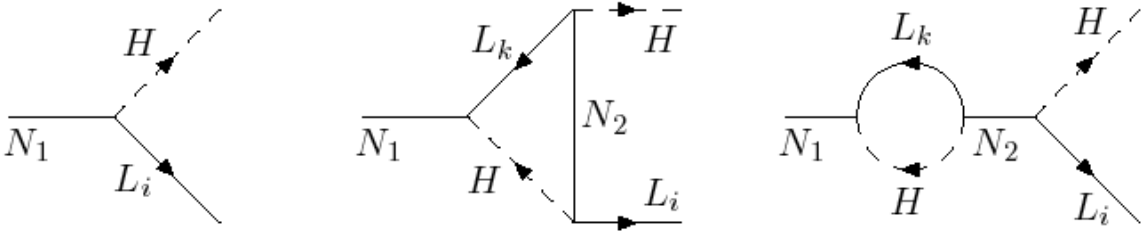}
\caption{The Feynman diagrams for $N_1$ decays in the usual leptogenesis scenario.
Only one heavier RH neutrino $N_2$ is displayed, but in general one sums over the same one-loop diagrams for each RH neutrino heavier than $N_1$.} \label{LHlepto}
\end{figure*}

Let us give a semi-quantitative description of leptogenesis. 
In general, the interactions of the lightest RH neutrino $N_1$ erase any asymmetry created at scales larger than its mass. The lepton asymmetry from $N_1$-decays is defined by
\begin{equation}
\epsilon_L = \frac{\Gamma(N_1\rightarrow LH) -\Gamma(N_1\rightarrow \bar{L}\bar{H})}{\Gamma(N_1\rightarrow LH)+ \Gamma(N_1\rightarrow \bar{L}\bar{H})}
~,
\end{equation} 
where $L$ and $H$ are the SM lepton and Higgs doublets.
The asymmetry arises from the interference between tree level and one loop Feynman diagrams for the decays of $N_1$, which are shown in Fig.\ref{LHlepto}.
Cosmological data determine the baryon number density over entropy ratio, $n_B/s \approx10^{-10}$. In the case of thermal leptogenesis one finds $n_B/s \approx - 10^{-3} \eta \epsilon_L$, where the washout factor $\eta$ is at most of order one  (see e.g. \cite{BDP1}). As a consequence, a sufficient baryon asymmetry can only be produced for 
$\epsilon_L\gtrsim 10^{-6}-10^{-7}$. 
Indicating with $Y_{Ni}$ the typical size of $N_i$ Yukawa couplings to lepton doublets and calling $M_i$ the mass of $N_i$,
a rough estimate gives
\begin{equation}
\epsilon_L \sim \sum_{i\ne 1} Y_{Ni}^2 \frac{M_1}{M_i} ~.
\label{eps}\end{equation}
A very similar combination of parameters determines light neutrino masses through the seesaw mechanism:
\begin{equation}
\frac{m_\nu}{v} \sim \sum_i Y_{Ni}^2 \frac{v}{M_i} ~.
\label{mnu}\end{equation}
Since $m_\nu/v\sim 10^{-12}$, the comparison of eqs.(\ref{eps}) and (\ref{mnu}) and the requirement $\epsilon_L \sim 10^{-6}$ implies $M_1\gtrsim 10^6 v \sim 10^8$ GeV \cite{DIM1}.
This means that leptogenesis may be successful only if it takes place at super-heavy scales.
Qualitatively the same lower bound on the leptogenesis scale persists  \cite{hamsen} in models including the contribution to $m_\nu$ from $SU(2)_L$ triplet scalars (type II seesaw mechanism). 
Even a fine-tuned cancellation between the type I and II seesaw
contributions to $m_\nu$ does not help, since the relation $\epsilon_L \propto m_\nu
= m_\nu^I + m_\nu^{II}$ still holds \cite{AK}.

The need for a super-heavy scale is phenomenologically unpleasant, since it is not directly accessible experimentally.
Moreover, 
in most (unified) extensions of the SM, the neutrino Yukawa couplings $Y_N$ are related to those of charged fermions, therefore one expects $Y_N$ to have a hierarchical structure. This implies a further suppression of $\epsilon_L$, which may prevent successful leptogenesis even at super-heavy scales. 
Finally, a reheating temperature $T_{RH}$ larger than $M_1$ is needed in order for thermal leptogenesis to occur.
If one invokes supersymmetry above the scale $v$, in particular to stabilize the large hierarchy $M/v$, then $T_{RH}\gtrsim 10^8$ GeV leads in general to overproduction of gravitinos, in severe tension with cosmological data (see \cite{KMY} and references therein).

The lower bound on the leptogenesis scale can be circumvented by some strong tuning of parameters: if
the two lightest RH neutrinos are quasi-degenerate in mass, eq.(\ref{eps}) does not hold and the asymmetry $\epsilon_L$ may be resonantly enhanced. As a consequence, the leptogenesis scale $\sim M_1 \approx M_2$ can be as small as $\sim$ TeV (see e.g. \cite{pilaU}). In this case the $Y_N$ couplings should be also tuned, since the contributions $\sim Y_{Ni}^2/M_i$ to $m_\nu$ need to cancel each other. In this talk we do not consider the resonant scenario, rather we develop a minimal alternative.

  
\section{LEPTON YUKAWA COUPLINGS IN STANDARD MODEL EXTENSIONS}

The non-zero neutrino masses and leptogenesis require to add several ingredients to the SM:
new sterile leptons, a super-heavy scale for their mass, Yukawa couplings between them and SM lepton doublets. One may wonder if a complete theory, which naturally accommodates these ingredients, does contain other sources of lepton asymmetry. If this were the case, some of the constraints  discussed in the previous section could be evaded.

Let us begin by asking what scalar multiplets couple a RH neutrino $N$ to the SM leptons $L$ and $l_R$. In full generality, the two Yukawa interactions read
\begin{equation}
-Y_N H \bar{N}L - Y_R \delta^+ N^T C l_R + {\rm h.c.}~,
\label{ynr}\end{equation}
where $H=(H^+~H^0)^T$ has the quantum numbers of the SM Higgs, while $\delta^+$ is a $SU(2)_L$ singlet with charge $+1$. In the usual leptogenesis scenario only the coupling $Y_N$ is considered. However, if $\delta^+$ is lighter than RH neutrinos,
the coupling $Y_R$ allows for the alternative decay channel 
$N \rightarrow \delta^+ l_{R}$, which may provide a new source of lepton asymmetry \cite{FHM}.
Notice that  only $SU(2)_L$ singlet particles are involved in the decay. 
In general, the charged singlet $\delta^+$ may couple also to SM left-handed leptons as
\begin{equation}
 - Y_L \delta^+ L^T C i\tau_2 L + {\rm h.c.}~,
\label{yl}\end{equation}
where the matrix $Y_L$  is antisymmetric.

Several simple extensions of the SM contain the singlet $\delta^+$:
\begin{itemize}
\item In $SU(5)$ models, $\delta^+$ is part of the Higgs multiplet $10_H$.
\item In left-right symmetric models, based on the gauge group $SU(2)_L\times SU(2)_R\times U(1)_{B-L}$, there are two possibilities: either $\delta^+ \sim (1,1,2)$ or $\delta^+$ can be identified with the singly charged component of an $SU(2)_R$ triplet $\Delta_R\sim(1,3,2)$.
\item In SO(10) models, which contain the left-right symmetric group, the two possibilities correspond to $\delta^+ \in 120_H$ or $\delta^+ \in \overline{126}_H$, respectively.
\end{itemize}
The coupling (\ref{yl}) is forbidden by the gauge symmetry in the case $\delta^+ \in \Delta_R$ ($\delta^+ \in\overline{126}_H$).

Let us consider the minimal left-right symmetric model.  The RH triplet $\Delta_R$ couples to RH lepton doublets $R=(N ~ l_R)^T$ as
\begin{equation}
-Y_\Delta R^T C i \tau_2 \Delta_R R + {\rm h.c.} ~.
\label{yd}\end{equation}
The vacuum expectation value $v_R$ of the $\Delta_R$ neutral component breaks $SU(2)_R$ and gives mass to the RH neutrinos. At the same time, the interaction (\ref{yd}) contains the second term in eq.(\ref{ynr}), so that $Y_\Delta$ and $Y_R$ are identified. The RH neutrino mass matrix $M_R = v_R Y_\Delta$ and the coupling matrix $Y_R = \sqrt{2} Y_\Delta$ are diagonalized at the same time, therefore there is no flavor mixing in the RH sector: the $i$-th RH neutrino $N_i$ decays only into $\delta^+ l_{Ri}$ and
no lepton asymmetry is generated. To achieve leptogenesis, one has to resort to less minimal models, e.g., (i) to introduce a second $SU(2)_R$ triplet; (ii) to admit a source of RH neutrino masses other than $v_R$; (iii) to add a $\delta^+$ which is singlet under both $SU(2)_L$ and $SU(2)_R$.


\section{RIGHT-HANDED SECTOR LEPTOGENESIS}

In this section we compute the lepton asymmetry generated in the decays of the lightest RH neutrino through the $Y_R$ coupling,
$N_1 \rightarrow \delta^+ l_{Ri}$. 
We assume $M_\delta < M_1$, where $M_\delta$ is the mass of $\delta^+$. In the opposite case a lepton asymmetry would be produced in $\delta^+$ decays, but it would be washed out by $N_i$ interactions as well as by gauge scattering (more details in \cite{FHM}).

\begin{figure*}[t]
\centering
\includegraphics[width=30mm,angle=270]{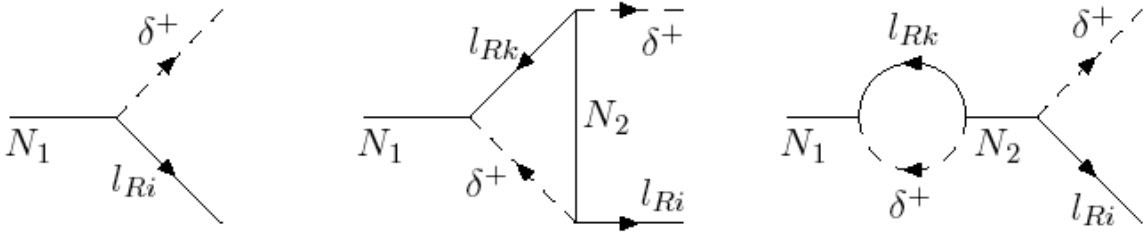}
\caption{The Feynman diagrams for $N_1$ decays in the right-handed sector leptogenesis scenario.} \label{RHlepto}
\end{figure*}

The decay of $N_1$ occurs through the same tree and one loop Feynman diagrams as the usual leptogenesis, by replacing $L_i$ with $l_{Ri}$ as well as $H$ with $\delta^+$,
as shown in Fig.\ref{RHlepto}. While $H$ has zero lepton number, the lepton number of $\delta^+$ has to be determined by its dominant decay channel. The interaction $Y_L$ given in eq.(\ref{yl}) leads to the two-body decay $\delta^+ \rightarrow l^+\bar{\nu}$, so that $L(\delta^+)= -2$. If $Y_L$ is forbidden, 
than $\delta^+$ may only decay into three bodies through a virtual $N$.
The dominant channel turns out to be $\delta^+ \rightarrow l^+ l^- H^+$ \cite{FHM}, so that $L(\delta^+)\approx 0$. In both cases, a net lepton number is produced in the decay $N_1\rightarrow \delta^+ l_{Ri}$ and therefore a lepton asymmetry may be generated.
We call this scenario RH sector leptogenesis.

Of course $N_1$ decays also through the usual Yukawa coupling $Y_N$. For simplicity, in the following we will not display this additional contribution to the lepton asymmetry. 
In fact, it is negligible as long as $M_1\ll 10^8$ GeV, as demonstrated in section 1.

We consider for definiteness only one RH neutrino heavier than $N_1$ (the results can be immediately generalized to an arbitrary number of RH neutrinos). The computation of the lepton asymmetry then gives
\begin{equation}
\epsilon_L=\frac{1}{8\pi} Y_{R2}^2 \frac{M_1}{M_2} ~,~~~~~
Y_{R2}^2 \equiv \frac{{\rm Im}(Y_R Y_R^\dag)_{12}^2}{\sum_i |(Y_R)_{1i}|^2} ~,
\label{epsR}\end{equation}
up to corrections of order $(M_\delta/M_1)^2$ and $(M_1/M_2)^2$.
The quantity $Y_{R2}$ is of the order of the Yukawa couplings of $N_2$ to $l_{Ri}$.
The $N_1$ decay width is given by
\begin{equation}
\Gamma_1=\frac{1}{16\pi} Y_{R1}^2 M_1 ~,~~~~ Y_{R1}^2 \equiv \sum_i |(Y_R)_{1i}|^2 ~,
\end{equation}
where $Y_{R1}$ is of the order of the Yukawa couplings of $N_1$ to $l_{Ri}$.
In order to suppress washout from $N_1$ inverse decays, one has to impose
the out-of equilibrium condition  
$\Gamma_1 \lesssim H(T=M_1)$, where $H$ is the Hubble constant. 
Requiring also $\epsilon_L \gtrsim 10^{-6}$, we find the constraint
\begin{equation}
\frac{Y_{R1}}{Y_{R2}} \lesssim 0.2 \cdot  \sqrt{\frac{M_{1}}{M_{2}}}\sqrt{\frac{M_{1}}
{10^9{\rm GeV}}} ~.
\end{equation}
Therefore, RH sector leptogenesis is successful even for $M_1$ much smaller than $10^9$ GeV, provided a hierarchy in the Yukawa coupling matrix $Y_R$ is assumed.
In particular, $M_1$ as light as few TeVs requires $Y_{R1}/Y_{R2} \sim 10^{-4}$, which is comparable with observed hierarchies in charged fermion Yukawa matrices.
We checked that other sources of asymmetry washout (in particular $\Delta L=2$ scattering due to $N_2$ exchange) are also under control down to these low scales
\cite{FHM}.

An illustrative set of parameters leading to successful leptogenesis is the following:
\begin{equation}\begin{array}{c}
M_{1}\simeq 2 {\rm TeV} ~,~~ M_{2}\simeq 6 {\rm TeV} ~,~~ M_{\delta}\simeq 750{\rm GeV} ~,\\
Y_{R2}\simeq 4 \cdot 10^{-3} ~,~~ Y_{R1}\simeq 10^{-7} ~.
\end{array}\end{equation} 
In fact, leptogenesis can be successful for $M_\delta$  as small as allowed by direct searches at collider ($\sim 100$ GeV). The lightest RH neutrino mass $M_1$ needs to be few times larger than $v$ to allow for lepton to baryon asymmetry conversion by sphalerons. The mass $M_2$ has to be heavier than $\sim 4$ TeV to suppress the washout induced by $N_2$ mediated $\Delta L=2$ scatterings
(for this estimate we did not consider the additional resonant enhancement of the asymmetry which may occur if $M_1\approx M_2$).

Let us restate the reason why the scale of RH sector leptogenesis can be much lower than in the usual leptogenesis scenario. In the present case eq.(\ref{eps}) is replaced by $\epsilon_L \sim Y_{R2}^2 M_1/M_2$
and $Y_{R2}$ can be taken relatively large without inducing any neutrino masses, which depend only on $Y_N$ as in eq.(\ref{mnu}). In other words, barring resonant effects, TeV scale RH neutrinos require very small $Y_N$ couplings and leptogenesis at low scale may occur only if another set of couplings ($Y_R$) is introduced in the theory. Contrary to the resonant leptogenesis scenario, no states quasi-degenerate in mass are needed, nor cancellations between $Y_N$ couplings. The price to pay is to introduce an extra degree of freedom, the scalar $\delta^+$.


\section{PHENOMENOLOGICAL SIGNATURES AT TeV SCALE}

Leptogenesis relies on the assumption that heavy RH neutrinos exist. In the usual scenario, they decay into a lepton doublet and the SM Higgs doublet, yet to be discovered at LHC. A sufficiently light charged singlet scalar $\delta^+$, if any, would also be discovered in colliders, by Drell-Yan pair production: $q\bar{q}~(e^+ e^-) \rightarrow \gamma ~(Z) \rightarrow  \delta^+ \delta^-$ (for an estimate of the cross-section see \cite{FHM}). If $\delta^+$ were observed, RH leptogenesis would become as plausible as the usual one, since  RH neutrinos will naturally decay into $\delta^+ l_{R}$.

If $\delta^+$
couples (antisymmetrically) to left-handed leptons as in eq.(\ref{yl}), 
it would mainly decay
into antilepton and antineutrino. The decay into $\tau^+$ 
is the more
interesting one, because it can be used to identify at the LHC 
a relatively light MSSM charged Higgs $H^+$ \cite{hashemi}.  
The two particle decays can
be distinguished by analyzing the angular distribution of the outgoing
antilepton, since it is left-handed in the case of $H^+$ and right-handed
in the case of $\delta^+$. 
If the  coupling in eq.(\ref{yl}) is absent, $\delta^+$ can 
decay only very slowly to three bodies.
Therefore,  the $\delta^+\delta^-$ pair will leave 
in the detector a pair of long curved charged particle tracks which could 
be distinguished from a muon pair by the fact that they are 
less relativistic.

The $\delta^+$ exchange may also mediate lepton number violating processes. In the case
of $\mu\rightarrow e\gamma$, the dominant contribution is found to come from the $Y_L$ couplings and can be as large as the present experimental limit if $M_\delta \lesssim 1$ TeV
\cite{FHM}.

If also $N_2$ is as light as few TeVs, it can be produced through the relatively large Yukawa coupling $Y_{R2} > 10^{-3}$. Notice that such low scales RH neutrinos in general erase any lepton asymmetry produced at higher temperatures. Therefore their observation would indicate the need for leptogenesis at low scale, as allowed by our model.

\section{SUMMARY}

The usual scenario of baryogenesis via leptogenesis relates the matter-antimatter asymmetry to neutrino masses, but there is a price to pay in order to maintain such relation: 
a sufficient lepton asymmetry is produced only at super-heavy scale, larger than $10^8$ GeV.

On the contrary, we have shown that, if RH neutrinos decay into RH charged leptons, then leptogenesis may work at scales as low as $\sim$ TeV. 
This requires to add to the SM with RH neutrinos one charged scalar $\delta^+$, singlet under $SU(2)_L$. Such particle can be easily incorporated in simple (unified) extensions of the SM. We called this scenario RH sector leptogenesis.

A sufficiently light $\delta^+$ (at or below the TeV scale) can be easily pair-produced at colliders and its decays have characteristic signatures. It may also mediate sizable lepton flavor violation processes like $\mu\rightarrow e\gamma$. While the lightest RH neutrino $N_1$
is very weakly coupled and therefore unobservable, $N_2$ can be also as light as few TeVs and may be produced through the same sizable coupling $Y_{R2}$ that generates the lepton asymmetry.

\bigskip 
\begin{acknowledgments}
I would like to thank Yasaman Farzan for invitation and the Organizers
for local financial support. The efforts of them all made 
this week in Tehran a memorable experience.
The results presented in this talk were derived in collaboration with Thomas Hambye
and Ernest Ma.
My work is partially supported by the RTN European Program MRTN-CT-2004-503369.
\end{acknowledgments}

\bigskip 


\end{document}